\documentclass{ws-ijmpc}

\usepackage{graphicx}

\begin{document}

\markboth{Nuno Crokidakis}
{The dynamics of early transoceanic voyages: A resource-coupled model of crew health and survival}

\catchline{}{}{}{}{}

\title{{The dynamics of early transoceanic voyages: A resource-coupled model of crew health and survival}}

\author{Nuno Crokidakis $^{*}$}

\address{
Instituto de F\'{\i}sica, \hspace{1mm} Universidade Federal Fluminense \\
 Niter\'oi - Rio de Janeiro, \hspace{1mm} Brazil \\ 
$^{*}$ nunocrokidakis@id.uff.br}

\maketitle

\begin{history}
\received{Day Month Year}
\revised{Day Month Year}
\end{history}

\begin{abstract}
\noindent
The great transoceanic voyages of the Age of Discovery imposed severe human costs on the crews that sustained them. Long travel durations, deteriorating sanitary conditions and finite food and water supplies created an environment in which disease, exhaustion and mortality became central features of maritime expansion. Despite extensive historical documentation of these hardships, quantitative approaches rarely describe the voyage itself as a dynamical system coupling human health and resource depletion. In this work, we propose a minimal compartmental model for crew dynamics during long-distance oceanic expeditions. The population is divided into healthy, ill and dead individuals, while an additional dynamical variable represents the effective level of provisions on board. Disease incidence depends nonlinearly on resource availability, capturing the increasing health risk associated with deteriorating food, water and sanitary conditions. The resulting system of coupled ordinary differential equations is analyzed in the transient regime relevant to finite-duration voyages. Rather than reconstructing specific historical expeditions, the model aims to identify generic mechanisms governing the interplay between provisioning, health deterioration, mortality, and travel duration. Numerical results across a broad range of voyage durations and initial provisioning levels show that resource limitations may substantially reduce the healthy and operational fraction of the crew even when the overall survival fraction remains relatively high, revealing a clear distinction between demographic survival and operational viability. Increasing the initial provisioning level delays the crossover at which ill individuals become as numerous as healthy individuals and extends the period over which the crew remains predominantly operational. We further introduce a critical provisioning time, defined as the time required for resources to reach a prescribed stress threshold, and obtain an approximately linear dependence on the initial resource buffer. The framework provides a simple quantitative perspective on the internal dynamics of early transoceanic navigation and illustrates how dynamical-systems approaches can contribute to the study of historical processes.


\end{abstract}

\ccode{PACS Nos.: 05.10.-a, 87.23.Cc, 87.23.Ge}

\section{Introduction}

The Age of Discovery fundamentally depended on long transoceanic voyages that connected Europe to Africa, the Americas and Asia, profoundly reshaping global trade, political relations and cultural exchange \cite{Diffie1977,Boxer1969}. These expeditions established new maritime routes and permanently altered the course of world history. Historical narratives have therefore traditionally emphasized the preparation of expeditions, the arrival at new territories, and their geographical, political and economic consequences \cite{Diffie1977,Boxer1969}. By contrast, the voyages themselves are often treated primarily as the interval separating departure from arrival, despite representing the stage during which the physical condition of an expedition could change dramatically. In other words, while history has extensively documented where these voyages began and where they ended, comparatively less attention has been devoted to the internal dynamics of what happened in between.

Life aboard early modern ships was characterized by severe environmental and logistical constraints. Crews remained confined for weeks or months while relying on finite stocks of food and fresh water whose quantity and quality deteriorated during prolonged voyages \cite{Riley,Steckel}. Poor sanitary conditions, overcrowding, nutritional deficiencies, infectious diseases, and physical exhaustion constituted persistent threats \cite{Haines}. Scurvy, in particular, became one of the major health problems associated with extended oceanic navigation \cite{Carnemolla,Kinlin,Brown}. Historical records describe episodes of severe morbidity and mortality, illustrating the substantial human cost associated with early transoceanic expansion \cite{Riley,Worden}. Quantitative studies have employed demographic, statistical, and econometric approaches to reconstruct mortality patterns in maritime populations \cite{Riley,Steckel,Motesharrei}. Nevertheless, despite this extensive literature, comparatively little attention has been devoted to representing the internal evolution of a voyage as a coupled dynamical system, in which crew health and resource availability continuously affect one another throughout the crossing.

Importantly, these risks were not restricted to ordinary sailors. Officers and commanders could also become seriously ill or die during long-distance expeditions. Paulo da Gama, who commanded the S\~ao Rafael during Vasco da Gama's first voyage to India, became gravely ill during the return journey and died in the Azores before completing the voyage \cite{Ravenstein1}. Another remarkable example is Bartolomeu Dias, one of the most experienced Portuguese navigators of the period and the first European to round the Cape of Good Hope, who died in 1500 when the ship he commanded in Cabral's expedition was lost in a storm near the same cape \cite{Fonseca}. These episodes illustrate that, although strong social and hierarchical differences existed aboard ship, experience and rank did not eliminate the severe risks associated with early transoceanic navigation.

More broadly, mathematical approaches have increasingly been used to investigate historical processes through the language of dynamical and complex systems. This perspective is closely related to the development of historical dynamics and cliodynamics, in which interacting demographic, economic, political, and social variables are represented through explicit mathematical mechanisms in order to investigate the processes underlying historical change \cite{Turchin,Korotayev,Grinin}. Mathematical models have consequently been applied to phenomena ranging from long-term population and political cycles to the expansion and collapse of states and empires \cite{Turchin,Korotayev,Grinin}. For example, G\"und\"uz developed dynamical descriptions of the rise and fall of empires using concepts from nonlinear dynamics and statistical physics \cite{Gunduz}. More recently, minimal dynamical models have also been employed to investigate specific historical episodes, including the siege and fall of Syracuse \cite{Syracuse} and the collapse of the Inca empire following the Spanish arrival \cite{Inca_empire}. These approaches illustrate how simplified mathematical models can complement historical analysis by isolating mechanisms and feedbacks that may be difficult to identify from narrative descriptions alone. In a broader context, a similar modeling philosophy has been extensively employed in sociophysics, where concepts and methods from statistical physics and dynamical systems are used to construct minimal descriptions of complex collective phenomena in human societies \cite{Galam2008,Galam_book,Sooknanan,Castellano2009,CSF}. Such approaches emphasize that simplified models need not reproduce all microscopic details of a social system to provide insight into the mechanisms responsible for its macroscopic behavior.

From this perspective, a transoceanic voyage provides a particularly natural example of a coupled resource-population system. Crew members continuously consume finite provisions, while the progressive depletion and deterioration of these resources modify the incidence of disease and, ultimately, mortality. At the same time, changes in the health composition of the crew alter resource consumption, generating a feedback between human and logistical variables. The voyage can therefore be regarded not merely as transportation between two geographical points, but as a finite-time dynamical process in which resources and population health coevolve.

The present work adopts this perspective. Rather than reconstructing specific historical expeditions, such as those led by Pedro \'Alvares Cabral, Vasco da Gama or Christopher Columbus, our objective is to identify generic mechanisms governing crew health during prolonged transoceanic voyages. We propose a minimal resource-coupled compartmental model describing the evolution of healthy, ill and dead crew members together with an effective provisioning variable. The resulting system of ordinary differential equations is analyzed in the transient regime relevant to finite-duration voyages, allowing us to investigate how voyage duration and initial provisioning jointly shape the health, survival and operational condition of the crew.

The proposed model is intentionally minimal. Its purpose is not to reproduce particular historical voyages or estimate historical mortality rates, but rather to identify generic dynamical mechanisms that may have operated across a broad class of early maritime expeditions. In this sense, historical voyages provide the physical motivation for the model, whereas our emphasis lies on the collective behavior emerging from the interplay between finite resources, health deterioration and mortality in confined human populations. As we show below, this perspective reveals an important distinction between demographic survival and operational viability, since a substantial fraction of a crew may survive a voyage even after its healthy and operational component has been strongly depleted.


\section{Mathematical model}

\qquad We consider a single transoceanic voyage as a closed system consisting of a fixed number $N$ of crew members. At any time $t$, individuals are classified into three health states: Healthy, Ill and Dead. We denote by $H(t), I(t)$ and $D(t)$ the corresponding populations, satisfying the normalization condition $H(t) + I(t) + D(t) = N$, with death treated as an absorbing state.

To eliminate the dependence on the system size, we introduce the normalized variables $h(t) = \frac{H(t)}{N}, i(t) = \frac{I(t)}{N}$ and $d(t) = \frac{D(t)}{N}$, which obey the normalization condition 
\begin{equation}
h(t) + i(t) + d(t) = 1.
\end{equation}

In addition to the population variables, we introduce a dynamical quantity $P(t)$ representing the effective stock of provisions on board, including both quantity and quality of food and water, as well as broader sanitary conditions. We define the per-capita provisioning level as $p(t) = \frac{P(t)}{N}$, where $p(t)$ is expressed in units of characteristic per-capita provisioning scale and is therefore dimensionless.

The model contains a set of phenomenological parameters associated with disease propagation, mortality, recovery, consumption, and provisioning deterioration. The parameter $\alpha$ represents a baseline illness incidence unrelated to provisioning conditions, while $\beta$ controls the increase in disease risk induced by resource depletion. The parameter $\mu$ denotes direct mortality among otherwise healthy individuals, accounting for accidents and acute events during the voyage. Ill individuals recover with rate $\gamma$ and die with rate $\delta$. Finally, $c_1$ and $c_2$ represent the effective per capita consumption rates of provisions by healthy and ill crew members, respectively, whereas $\rho$ accounts for the spontaneous deterioration of provisions over time.

The central assumption of the model is that the incidence of illness depends on the provisioning level. The per capita rate at which healthy individuals become ill is given by
\begin{equation}
\lambda(p) = \alpha + \frac{\beta}{1+p},
\end{equation}
\noindent
where the second term captures the additional health risk associated with deteriorating provisions. This functional form is monotonic and saturating: for large $p$, the provisioning effect becomes negligible, whereas for $p \to 0$, the incidence approaches its maximum value $\alpha + \beta$.

The dynamics of the health compartments are described by the system
\begin{eqnarray} \label{eq1}
\frac{dh}{dt} & = & -\lambda(p)\,h - \mu\,h + \gamma\,i, \\ \label{eq2}
\frac{di}{dt} & = & \lambda(p)\,h - (\gamma + \delta)\,i, \\ \label{eq3}
\frac{dd}{dt} & = & \mu\,h + \delta\,i,
\end{eqnarray}
\noindent
while the provisioning dynamics obey
\begin{equation} \label{eq4}
\frac{dp}{dt} = -c_1 h - c_2 i - \rho p.
\end{equation}

Since $p(t)$ represents an effective provisioning level, its physical domain is restricted to $p \geq 0$. If complete depletion is reached, $p$ is kept at zero, thereby excluding unphysical negative provisioning levels.

\begin{figure}[t]
\begin{center}
\vspace{6mm}
\includegraphics[width=0.7\textwidth,angle=0]{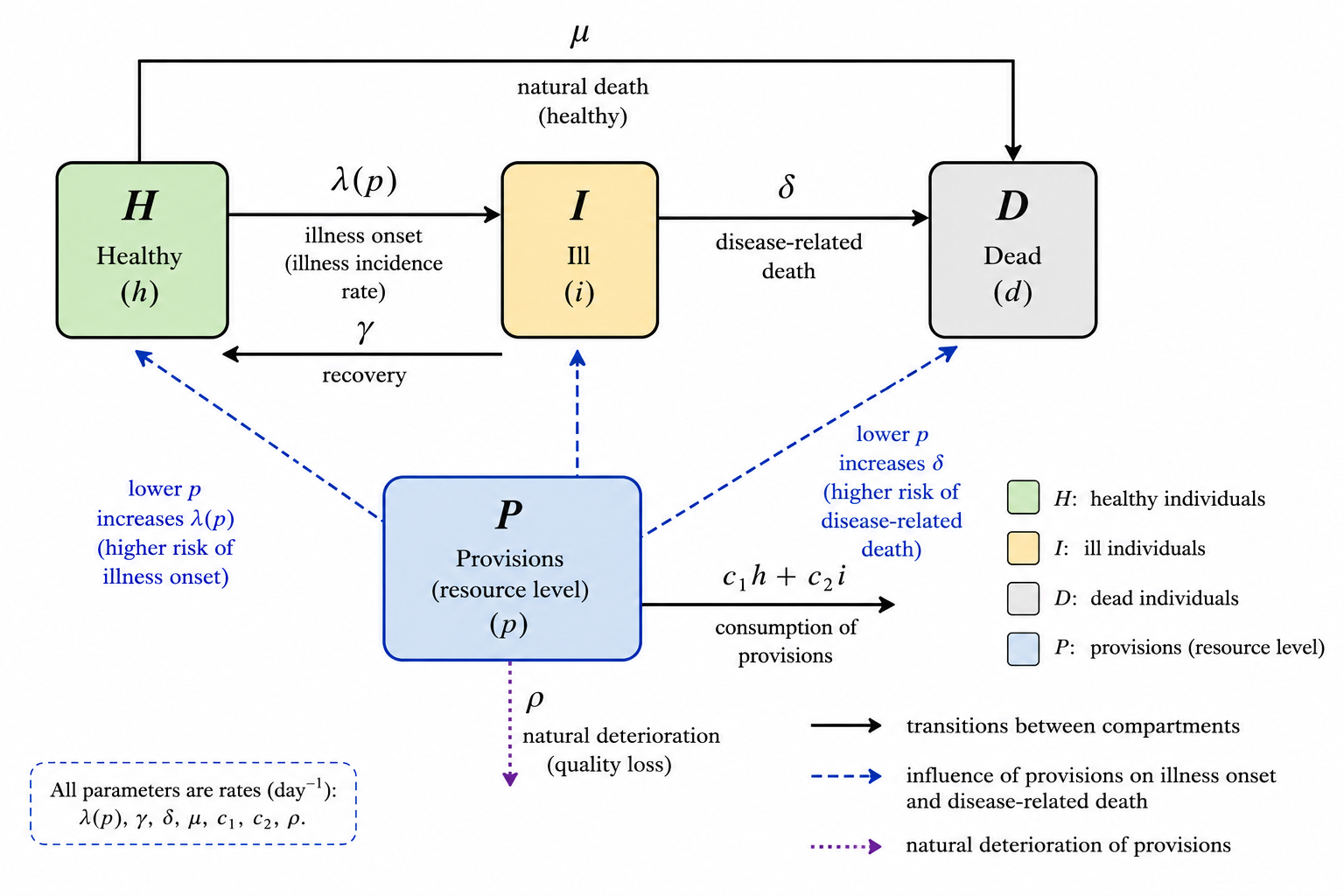}
\end{center}
\caption{Schematic representation of the model. Solid arrows indicate transitions between the Healthy ($H$), Ill ($I$) and Dead ($D$) compartments, while dashed arrows represent the coupling between crew health and the provisioning level $P$. The dotted arrow indicates the natural deterioration of provisions due to quality loss, independent of the crew health. All parameters are $day^{-1}$: $\lambda(p), \mu, \gamma, \delta, c_1, c_2$ and $\rho$.}
\label{fig1}
\end{figure}

A schematic illustration of the transitions among the compartments and their coupling to the provisioning dynamics is shown in Fig. \ref{fig1}. With this normalization, all parameters $\alpha, \beta, \mu, \gamma, \delta, c_1, c_2$ and $\rho$ are expressed in units of day$^{-1}$, whereas $h, i, d$ and $p$  are dimensionless quantities.

The present model is intentionally minimal and is designed to capture generic mechanisms rather than reproduce specific historical expeditions in quantitative detail. Several simplifying assumptions are therefore adopted. First, the crew is treated as a homogeneous population, neglecting differences between officers, sailors, soldiers or other social groups that may have experienced distinct living conditions and mortality risks.

This is, admittedly, a simplifying assumption. Historical accounts indicate that officers generally had better living conditions than ordinary sailors, including preferential access to food, fresh water and wine, as well as more comfortable accommodations. Consequently, different social groups aboard ship may have experienced distinct risks of disease and mortality. Nevertheless, the present work deliberately adopts a homogeneous crew in order to isolate the fundamental mechanisms arising from the coupling between resource depletion and crew health. Moreover, despite these hierarchical differences, historical evidence shows that prolonged voyages exposed the entire crew, including officers and captains, to severe environmental stress, and deaths among commanding officers were not uncommon. Introducing social stratification constitutes a natural extension of the present model and will be considered in future work. In this sense, the homogeneous approximation should be interpreted as the first-order description of the collective dynamics aboard the ship, rather than as a literal representation of its social organization. Second, the model assumes a closed system during the voyage, with no recruitment or external replenishment except in possible extensions involving re-supply stops.

The provisioning variable $p(t)$ should be interpreted as an effective quantity combining food availability, water quality, and broader sanitary conditions. As a consequence, the model does not distinguish explicitly between different diseases or nutritional deficiencies, such as scurvy, dehydration or infectious outbreaks. Instead, these effects are incorporated phenomenologically into the illness incidence rate $\lambda(p)$.

Another simplification concerns the deterministic nature of the dynamics. Random events such as storms, ship damage or abrupt epidemic outbreaks are not explicitly included. Likewise, the model does not attempt to describe geopolitical interactions occurring upon arrival at the destination, focusing exclusively on the internal evolution of the expedition during the crossing itself.

Despite these limitations, the model retains the essential feedback structure linking resource depletion, health deterioration and mortality in a confined population. In this sense, the framework should be viewed as a conceptual baseline capable of identifying robust qualitative regimes and threshold-like behaviors that may characterize a broad class of early transoceanic voyages.

The model describes the coupled evolution of crew health and provisioning during a transoceanic voyage, highlighting how resource depletion can drive nonlinear deterioration and threshold-like survival outcomes.

In the next section we will present our results.


\section{Results}

\qquad Unlike many population-dynamical models, the physically relevant regime here is intrinsically transient, since the voyage duration provides a natural finite observation time. The asymptotic stationary state is trivial, corresponding to complete depletion of provisions and eventual extinction of the crew. Therefore, the physically relevant regime is the transient dynamics over finite travel durations.

Thus, a voyage is described by integrating the system from an initial condition at $t=0$, corresponding to departure, up to a fixed travel duration $T$. We consider the initial conditions as $h(0)=1, i(0)=d(0)=0$, corresponding to a crew that is initially healthy and operational. In addition, we introduce the survival fraction at arrival,
\begin{equation} \label{eq_S}
S(T) = h(T) + i(T) = 1 - d(T),
\end{equation}
\noindent
which represents the fraction of the initial crew that remains alive at the end of the voyage. Notice that survival does not necessarily imply full operational capacity, since surviving individuals may belong either to the healthy or to the ill compartment. The model therefore allows us to distinguish between demographic survival and operational capacity, while simultaneously accounting for their coupling to the availability of provisions. This distinction makes it possible to describe voyages in which mortality remains relatively limited while the effective operational capacity of the crew undergoes substantial deterioration.

\begin{figure}[t]
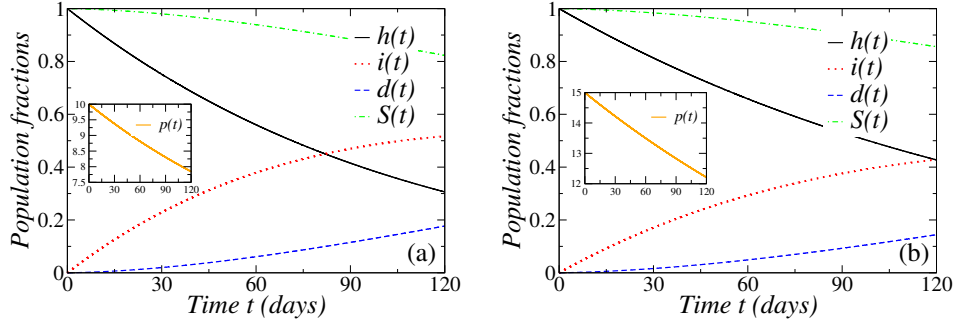

\begin{center}
\vspace{6mm}
\includegraphics[width=0.47\textwidth,angle=0]{figure2a.eps}
\hspace{0.3cm}
\includegraphics[width=0.47\textwidth,angle=0]{figure2b.eps}
\end{center}
\caption{Time evolution of the population fractions $h(t), i(t)$ and $d(t)$, together with the survival fraction $S(t)$, defined in Eq. \eqref{eq_S}, for a voyage duration $t_{max} = T = 120$ days and initial provisions $p_0=10$ (panel (a)) and $p_0=15$ (panel (b)). The insets show the corresponding evolution of the provision level $p(t)$. Increasing the initial provisioning level delays the operational crossover, defined by $h(t_{\times})=i(t_{\times})$, from $t_{\times}\approx 80$ days for $p_0=10$ to $t_{\times}\approx 120$ days for $p_0=15$. The fixed parameters are $\alpha=0.001, \beta = 0.09, \mu = 0.0002, \gamma = 0.0007, \delta = 0.004, c_1 = 0.012, c_2 = 0.006$ and $\rho=0.001$.}
\label{fig2}
\end{figure}

A characteristic timescale can be introduced from the internal composition of the surviving crew. We define the operational crossover time, $t_{\times}$, as the time at which the healthy/operational and ill fractions become equal, $h(t_{\times})=i(t_{\times})$. For $t<t_{\times}$, healthy/operational individuals constitute the largest fraction of the surviving crew, whereas for $t>t_{\times}$, ill individuals outnumber those who remain healthy and operational. Thus, $t_{\times}$ provides a simple measure of how long the crew can maintain a predominantly operational composition during the voyage. Importantly, this crossover does not imply population collapse or high mortality: the survival fraction $S(t)$ may remain relatively large even after the healthy fraction has been substantially depleted.

Fig. \ref{fig2} exhibits the time evolution of the population fractions $h(t), i(t)$ and $d(t)$, together with the survival fraction $S(t)$, defined in Eq. \eqref{eq_S}, for a voyage duration $t_{max} = T = 120$ days. Results are shown for initial  provisioning levels $p_0=10$ [panel (a)] and $p_0=15$ [panel (b)]. In both cases, the healthy and operational fraction $h(t)$ decreases continuously during the voyage, while the fractions of ill and dead individuals progressively increase. Remarkably, however, the survival fraction $S(t)$ decreases much more slowly than the healthy fraction. Thus, a substantial deterioration of the operational condition of the crew may occur even when the majority of individuals remain alive. This distinction between survival and operational capacity is an important feature of the dynamics.

Increasing the initial provisioning level from $p_0=10$ to $p_0=15$ substantially mitigates this deterioration. At a given time, the fraction of healthy individuals is larger, whereas both the ill and dead fractions are reduced. This effect is particularly evident in the operational crossover time $t_{\times}$, that was above defined by $h(t_{\times})=i(t_{\times})$. For $p_0=10$, the crossover occurs at approximately $t_{\times} \simeq 80$ days, whereas for $p_0=15$ it is delayed to approximately $t_{\times} \simeq 120$ days. Therefore, increasing the initial provisioning level does not merely improve survival; it also extends the period during which healthy individuals constitute the largest fraction of the crew. At the same time, $S(t)$ remains comparatively high in both cases, reinforcing the distinction between demographic survival and operational viability. The origin and broader implications of this behavior will become clearer in the following analysis.

An additional characteristic timescale can also be introduced to quantify the onset of resource stress during the voyage. We define the critical provisioning time, $t_c$, as the time at which the available provisions reach a prescribed critical level $p_c$, such that
\begin{equation} \label{eq_pc}
p(t_c) = p_c ~.
\end{equation}
The threshold $p_c$ represents a minimum provisioning level below which the voyage is considered to enter a regime of severe resource stress. Accordingly, the condition $t_c > T$ indicates that the provision level remains above $p_c$ throughout the entire voyage, whereas $t_c < T$ implies that the critical provisioning regime is reached before the end of the voyage. This definition is more general than requiring complete resource exhaustion and accounts for the fact that severe logistical stress may arise well before the available provisions vanish completely.

The relation between $t_c$ and the initial provisioning level $p_0=p(0)$ can be obtained numerically. However, a simple analytical approximation can be also be derived from Eq. \eqref{eq4}. To this end, let us neglect the spontaneous deterioration of provisions ($\rho=0$), as well as significant variations in the fractions of healthy $h$ and ill $i$ individuals during the voyage. Under these assumptions, Eq. \eqref{eq4} reduces approximately to $dp/dt \approx -c$, where $c$ represents an effective constant consumption rate. Direct integration yields $p(t) \approx p_0 - c\,t$. Imposing the critical condition $p(t_c)=p_c$, we obtain
\begin{equation} \label{eq_tc}
t_c \approx \frac{p_0-p_c}{c}.
\end{equation}
\noindent
Thus, for a fixed critical provisioning level $p_c$, the time required to reach the resource-stress threshold increases linearly with the excess initial provisions $p_0 - p_c$. Equivalently, imposing $t_c=T$ gives
\begin{equation} \label{eq_cT}
p_0 \approx p_c + c\,T ~,
\end{equation}
\noindent
showing that the minimum initial provisioning required to remain above the critical level throughout a voyage of duration $T$ increases approximately linearly with the travel time.

Since the present model is not intended as a quantitative reconstruction of a particular historical expedition, $p_c$ should be interpreted as a phenomenological threshold rather than as a historically calibrated provisioning level. We therefore consider different values of $p_c$ to test the robustness of the predicted scaling behavior.

\begin{figure}[t]
\begin{center}
\vspace{6mm}
\includegraphics[width=0.6\textwidth,angle=0]{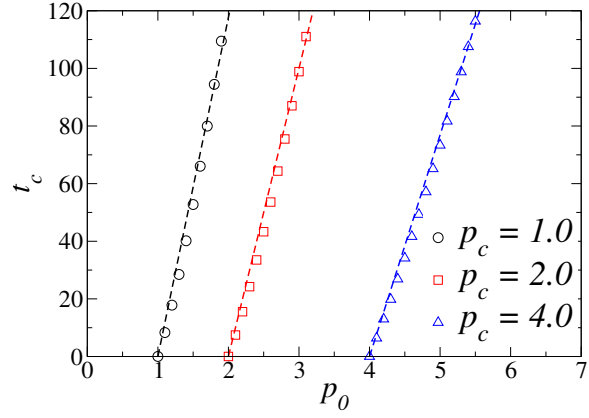}
\end{center}
\caption{Critical provisioning time $t_c$ as a function of the initial provisioning level $p_0$, for different values of the critical provisioning threshold $p_c$. Symbols represent numerical results, while dashed lines are guides to the eye following the analytical dependence predicted by Eq. \eqref{eq_tc}. The remaining fixed parameters are the same as those used in Fig. \ref{fig2}.}
\label{fig3}
\end{figure}

In Fig. \ref{fig3}, we show the numerically obtained values of $t_c$ for three distinct critical provisioning levels, namely $p_c=1.0, 2.0$ and $4.0$. As guides to the eye, we also plot straight lines of the form predicted by Eq. \eqref{eq_tc}, using different values of the effective constant $c$ for each value of $p_c$. No fitting procedure was performed, since Eq. \eqref{eq_tc} is intended as a heuristic approximation aimed at capturing the functional dependence of $t_c$ on the initial resource buffer $p_0 - p_c$, rather than providing a quantitative estimate of the effective consumption rate $c$. The values of $c$ shown in Fig. \ref{fig3} were therefore chosen only to illustrate the approximately linear behavior of the numerical results.

The numerical results closely follow the predicted linear dependence over the entire time window shown, $t_c \leq 120$ days, which corresponds to the maximum voyage duration considered throughout this work. Notice also that different effective values of $c$ are required for different values of $p_c$. This is expected, since $c$ summarizes the full time-dependent depletion dynamics into a single effective constant and is therefore not a fundamental parameter of the original model.

Importantly, the approximately linear behavior predicted by Eq. \eqref{eq_tc} is not a consequence of a particular choice of the critical provisioning level $p_c$. Instead, it persists when the criterion used to define the onset of logist stress is varied, suggesting that the scaling $t_c\sim (p_0 - p_c)$ is a robust feature of the model.

\begin{figure}[t]
\begin{center}
\vspace{6mm}
\includegraphics[width=0.47\textwidth,angle=0]{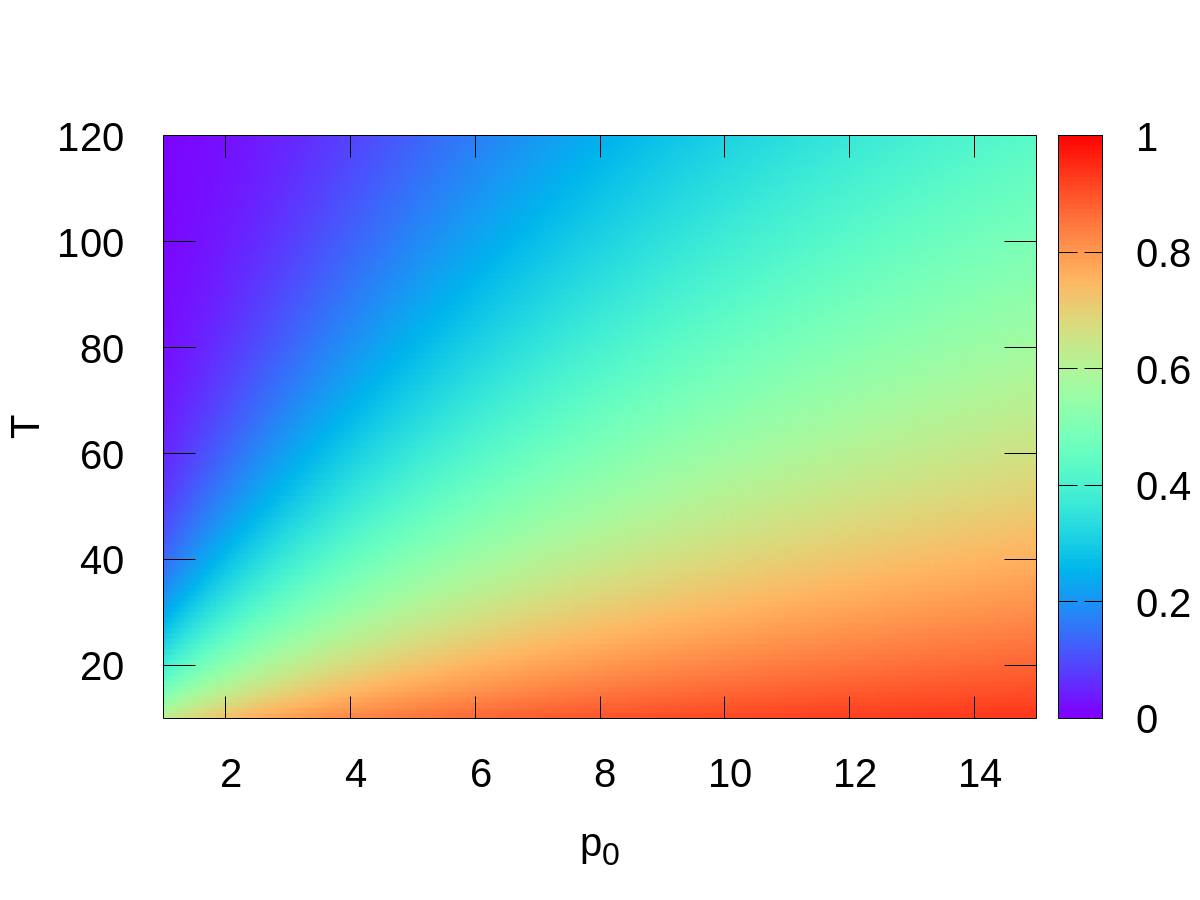}
\hspace{0.3cm}
\includegraphics[width=0.47\textwidth,angle=0]{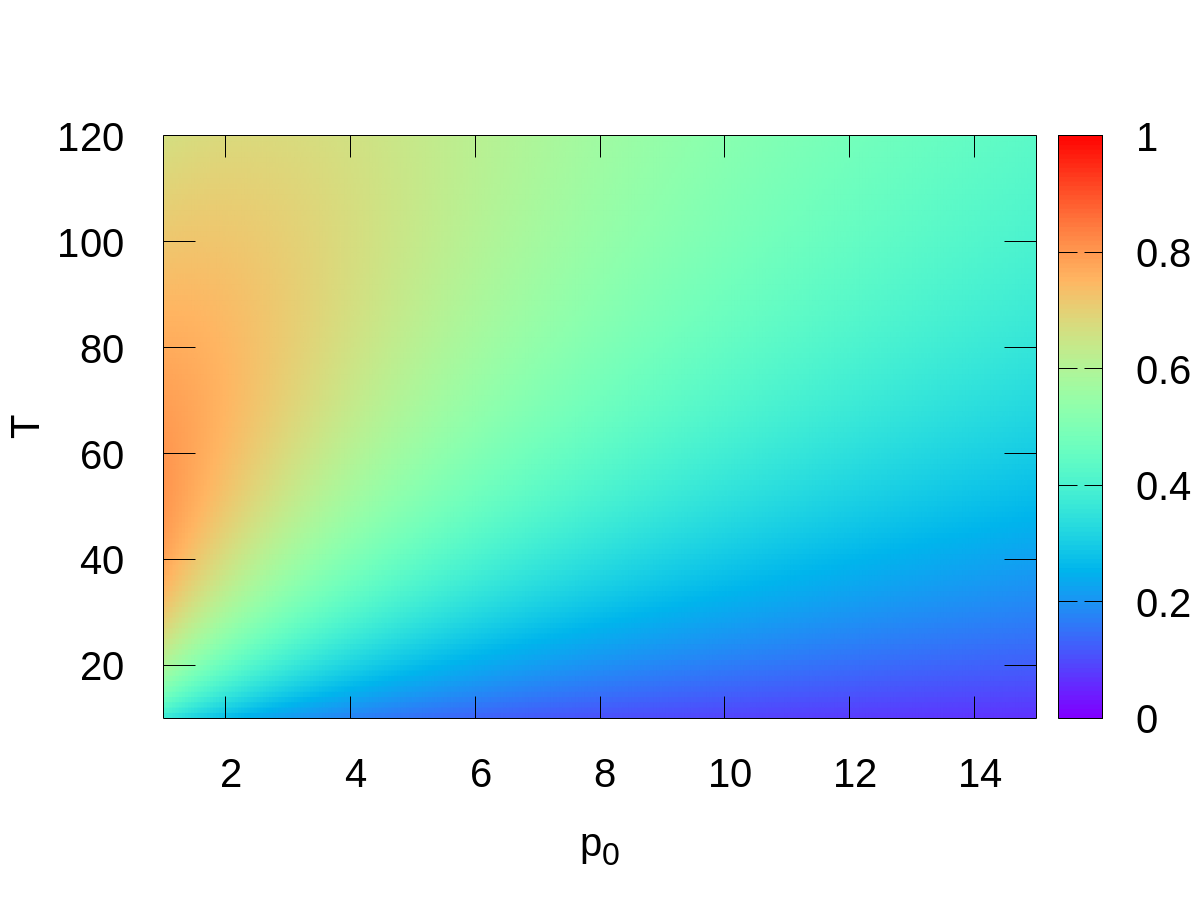}
\\
\vspace{0.2cm}
\includegraphics[width=0.47\textwidth,angle=0]{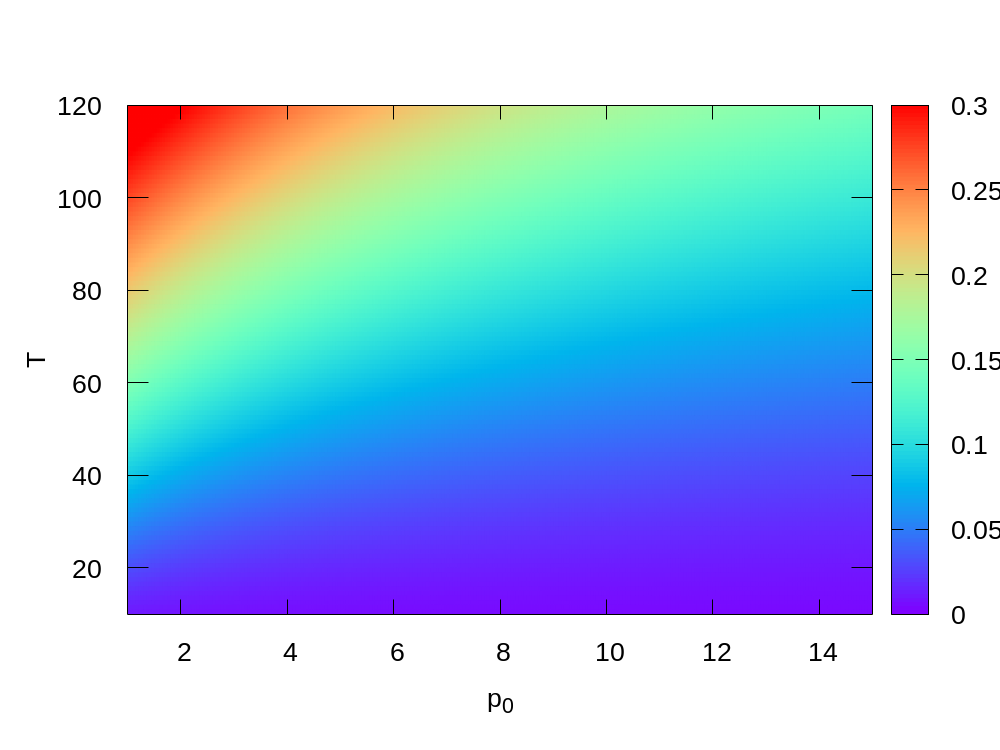}
\hspace{0.3cm}
\includegraphics[width=0.47\textwidth,angle=0]{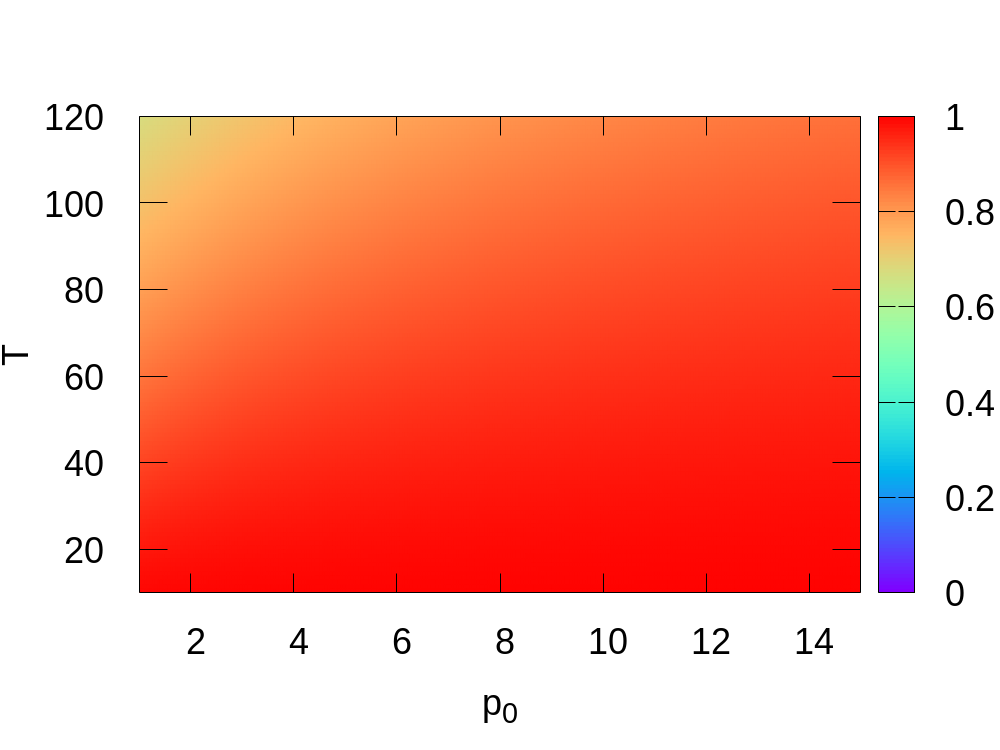}
\end{center}
\caption{Heatmaps of the final population fractions in the plane defined by the voyage duration $T$ and the initial provisions $p_0$. The color scale represents: (upper left) the healthy/operational fraction $h(T)$; (upper right) the ill fraction $i(T)$; (lower left) the dead fraction $d(T)$; and (lower right) the survive fraction $S(T)$. The remaining fixed parameters are the same as those used in Fig. \ref{fig2}.}
\label{fig4}
\end{figure}

In Fig. \ref{fig4} we show heatmaps for the subpopulations fractions $h(T), i(T)$ and $d(T)$, as well as for the survival density $S(T)$, in the plane $p_0$ vs $T$. The behavior observed in the time evolution of the subpopulation fractions becomes clearer when a broad range of initial provisioning levels and voyage durations is considered. For low values of $p_0$, a predominantly healthy crew can only be maintained for relatively short voyages (see the upper left panel of Fig. \ref{fig4}). As $T$ increases, the healthy fraction rapidly decreases, indicating a progressive loss of operational capacity. Increasing $p_0$, however, substantially extends the range of voyage durations over which a significant fraction of the crew remains healthy. For intermediate provisioning levels, relatively high values of $h$ persist up to considerably longer voyages, while for large $p_0$ the healthy fraction remains substantial even for durations approaching the longest voyages considered here. In particular, for sufficiently large $p_0$, more than half of the crew remain healthy even for $T$ close to $120$ days. Thus, increasing the initial provisioning level effectively shifts the deterioration of the crew toward progressively longer voyage durations.

The corresponding behavior of the ill fraction $i$ reinforces this interpretation (see the upper right panel of Fig. \ref{fig4}). Large concentrations of ill individuals occur predominantly in the region of long voyages combined with low initial provisioning. In contrast, increasing $p_0$ strongly suppresses the accumulation of illness, even when $T$ is large. The mortality map $d$ displays even clearer localization of adverse outcomes: significant mortality is concentrated mainly at long voyage durations and low provisioning levels, whereas increasing the initial resource stock produces a substantial reduction in the final dead fraction (see the lower left panel of Fig. \ref{fig4}). Taken together, the first three panels show that $p_0$ acts as a strong control parameter for the health composition of the crew, effectively compensating, to some extent, for the detrimental effect of increasing voyage duration $T$.

A particularly interesting picture emerges when these results are compared with the survival fraction $S$ (see the lower right panel of Fig. \ref{fig4}). In contrast to the pronounced variations observed in $h$ and $i$, survival remains relatively high over most of the $(p_0,T)$ plane. A substantial reduction in $S$ occurs mainly in the extreme region combining very low initial provisioning and long voyage duration. Even there, however, the model predicts an important distinction between mortality and loss of operational capacity. For instance, in the region of $p_0 \lesssim 2$ and $T \gtrsim 100$, approximately $70\%$ of the initial crew still be alive at arrival, while only a small fraction remains healthy and a large part of the surviving populations is ill. Therefore, a voyage that appears relatively successful when assessed solely through survival may correspond to a severely debilitated crew from an operational perspective. This result generalizes that behavior already observed in Fig. \ref{fig2}: demographic survival and operational viability represent distinct outcomes of the voyage, and the latter may deteriorate substantially before mortality becomes dominant. This apparent robustness of $S$ originates from its definition $S= h + i$: transitions from the healthy to the ill compartment strongly affect operational capacity while leaving the survival fraction unchanged.


\section{Final Remarks}

\qquad In this work, we proposed a minimal dynamical framework to investigate the internal evolution of crew health and resource availability during long transoceanic voyages. The crew was divided into healthy, ill and dead individuals, whose dynamics were coupled to a finite provisioning level representing the combined availability and quality of essential resources on board. Rather than attempting to reconstruct or quantitatively reproduce a particular historical expediction, the model was designed to identify generic mechanisms arising from the interaction between voyage duration, resource depletion, health deterioration and mortality. Accordingly, the numerical results should be interpreted as dynamical scenarios illustrating these mechanisms rather than as calibrated reconstructions of specific historical voyages.

The results reveal a clear distinction between demographic survival and operational viability. In particular, the survival fraction may remain relatively high even after the healthy and operational fraction of the crew has been substantially depleted. This occurs because deterioration of crew conditions initially transfers individuals from the healthy to the ill compartment without immediately affecting the total surviving population. Consequently, mortality alone may provide an incomplete characterization of the human impact of a long voyage: an expedition may reach its destination with most of its crew alive while a substantial fraction of the survivors is ill and potentially unable to perform normal duties. The heatmaps obtained over the $(p_0,T)$ plane show that this behavior is not restricted to particular trajectories, but persists over a broad range of initial provisioning levels $p_0$ and voyage durations $T$.

Initial provisioning plays a central role in controlling this deterioration. Increasing $p_0$ extends the range of voyage durations over which a predominantly healthy crew can be maintained and delays the operational crossover time $t_{\times}$, defined by $h(t_{\times})=i(t_{\times})$. A second characteristic timescale emerges from the resource dynamics. By defining the critical provisioning time $t_c$ through $p(t_c)=p_c$, where $p_c$ represents a prescribed resource-stress threshold, a simple analytical approximation yields $t_c \sim (p_0 - p_c)$. The numerical results reproduce this approximately linear dependence for different choices of $p_c$, suggesting that the initial resource buffer $p_0 - p_c$ provides a simple measure of how long an expedition can remain above a critical provisioning level. Together, $t_{\times}$ and $t_c$ characterize two complementary aspects of the voyage: the progressive degradation of the crew's operational condition and the approach to logistical stress.

A notable feature of the existing literature is the emphasis on the outcomes of early oceanic voyages, including discoveries, trade routes and geopolitical consequences, rather than on the internal dynamics of the voyages themselves. While historical accounts and medical studies document the hardships faced by crews, including disease, malnutrition and mortality, these processes are rarely framed as evolving dynamical systems. As a result, the voyage is often treated as a narrative interval connecting departure and arrival, rather than as a complex, time-dependent process in which resource depletion and health degradation interact. By focusing explicitly on this internal evolution, the present work shifts attention from outcomes to mechanisms and provides a quantitative framework for describing how the human condition of an expedition evolves during the crossing.

An important consequence of this perspective is that the model characterizes the internal state in which an expedition reaches its destination without attempting to predict what subsequently occurs there. The same expediction state, for example, one characterized by depleted provisions and a large fraction of ill individuals, may have very different consequences depending on the political, commercial and military environment encountered upon arrival. The Brazilian coast reached by the Portuguese expedition in 1500 and the highly competitive commercial environment encountered at Calicut, for example, represented very different external settings. Such differences lie outside the present model. Nevertheless, the framework suggests that the conditions accumulated during the crossing should not be regarded as irrelevant to the subsequent encounter: crews arriving with different levels of provisioning and operational capacity necessarily faced different constraints on their possible actions. In this sense, the voyage itself may constitute an important dynamical stage connecting the organization of an expedition at departure to the historical processes that followed its arrival.


The simplicity of the present framework also defines its limitations. The crew is treated as a homogeneous population, despite historical differences in access to food, water, accommodation and medical care among officers, sailors, soldiers and other groups. The provisioning variable combines several distinct resources and onboard sanitary conditions into a single effective quantity, and the model neglects stochastic events, weather conditions, navigation uncertainties, intermediate stops and resupply. These simplifications are deliberate, since the aim is to isolate a minimal resource-population feedback rather than reproduce individual voyages quantitatively. Future extensions could introduce heterogeneous crew classes, multiple resources, stochastic perturbations or explicit resupply events, allowing the same framework to address more detailed questions while preserving the basic dynamical mechanism considered here.

More broadly, the present results illustrate how methods from dynamical systems and complex systems modeling can complement traditional historical approaches by focusing on mechanisms that are difficult to isolate from narrative evidence alone. Long-distance voyages were not merely spatial displacements between historically important points: they were evolving systems in which finite resources and human health continuously interacted. Treating the crossing itself as a dynamical process provides a simple way to investigate these interactions and, more generally, to incorporate the dynamics of the journey into quantitative studies of early transoceanic expansion.


\section*{Acknowledgments}

The author acknowledges partial financial support from the Brazilian scientific funding agency Conselho Nacional de Desenvolvimento Cient\'ifico e Tecnol\'ogico (CNPq, Grants 308643/2023-2 and 406820/2025-2).

\bibliographystyle{elsarticle-num-names}

\begin{thebibliography}{00}


\bibitem{Diffie1977}
B. W. Diffie, G. D. Winius, \textit{Foundations of the Portuguese Empire, 1415-1580}, in Europe and the World in the Age of Expansion (University of Minnesota Press, 1977).


\bibitem{Boxer1969}
C. R. Boxer, \textit{The Portuguese seaborne empire, 1415-1825} (Hutchinson, London, 1969).


\bibitem{Riley}
J. C. Riley, \textit{Mortality on Long-Distance Voyages in the Eighteenth Century}, The Journal of Economic History 41(3), 651-656 (1981), doi:$10.1017/S0022050700044375$. 



\bibitem{Steckel}
R. H. Steckel, R. A. Jensen, \textit{New Evidence on the Causes of Slave and Crew Mortality in the Atlantic Slave Trade}, The Journal of Economic History 46(1), 57-77 (1986), doi:$10.1017/S0022050700045502$.


  
\bibitem{Haines}
R. Haines, J. McDonald, R. Shlomowtz, \textit{Mortality and Voyage Length in the Middle Passage Revisited}, Explorations in Economic History 38, 503-533 (2001).



\bibitem{Carnemolla}
S. E. Carnemolla, \textit{Death from scurvy on Vasco da Gama's first journey to India (1497-1499)}, Medicina nei Secoli 15, 615-630 (2003).


  
\bibitem{Kinlin}
L. M. Kinlin, M. Weinstein, \textit{Scurvy: old disease, new lessons}, Paediatrics and International Child Health 43(4), 83-94 (2023).
  

  
\bibitem{Brown}
S. R. Brown, \textit{Scurvy: How a surgeon, a mariner, and a gentleman solved the greatest medical mystery of the age of sail} (The History Press Ltd, 2001).


\bibitem{Worden}
N. Worden, \textit{Below the Line the Devil Reigns': Death and Dissent aboard a VOC Vessel} 61, 702-730 (2009).  


\bibitem{Motesharrei}
S. Motesharrei, J. Rivas, E. Kalnay, \textit{Human and nature dynamics (HANDY): Modeling inequality and use of resources in the collapse or sustainability of societies}, Ecological Economics 101, 90-102 (2014).


  
%


\bibitem{Ravenstein1}
\textit{A Journal of the First Voyage of Vasco da Gama, 1497–1499}, edited by E. G. Ravenstein (Cambridge University Press, 2010), doi:$10.1017/CBO9780511708480.002$.  


\bibitem{Fonseca}
L. A. Fonseca, \textit{O Essencial Sobre Bartolomeu Dias} (Imprensa Nacional - Casa da Moeda (Coleç\~ao ``O Essencial'', nº 31, Lisboa 1987).


\bibitem{Turchin}
P. Turchin, \textit{Historical Dynamics: Why States Rise and Fall} (Princeton University Press, 2003), $https://www.jstor.org/stable/j.ctt1wf4d45$.


\bibitem{Korotayev}
P. Turchin, A. Korotayev, L. Grinin, \textit{Why Do We Need Mathematical Models of Historical Processes}, in History $\&$ Mathematics: Historical Dynamics and Development of Complex Societies (2006).

  
\bibitem{Grinin}
L. Grinin, A. Korotayev, \textit{Modeling and Measuring Cycles, Processes, and Trends}, History $\&$ Mathematics: Trends and Cycles (2014).


\bibitem{Gunduz}
G. G\"und\"uz, \textit{The dynamics of the rise and fall of empires}, International Journal of Modern Physics C 27, 1650123 (2016).
  

\bibitem{Syracuse}
N. Crokidakis, \textit{Modeling the Siege of Syracuse: Resources, strategy, and collapse}, International Journal of Modern Physics C 37, 2550127 (2026).



\bibitem{Inca_empire}
N. Crokidakis, \textit{Modeling the Fall of the Inca Empire: A Lotka-Volterra Approach to the Spanish Conquest}, Physics 8(1), 7 (2026).


\bibitem{Galam2008}
S. Galam, \textit{Sociophysics: A Review of Galam Models}, International Journal of Modern Physics C 19, 409-440 (2008).


\bibitem{Galam_book}
S. Galam, \textit{Sociophysics: A Physicist's Modeling of Psycho-political Phenomena} (Springer, Berlin, 2012).

  

\bibitem{Sooknanan}
J. Sooknanan, D. M. G. Comissiong, \textit{Beyond peer pressure: Re-examining transmission dynamics in compartmental models of socially transmitted ills},  International Journal of Modern Physics C 2750121 (2027), doi:$10.1142/S012918312750121X$.




\bibitem{Castellano2009}
C. Castellano, S. Fortunato, V. Loreto, \textit{Statistical physics of social dynamics}, Reviews of Modern Physics 81, 591 (2009).


\bibitem{CSF}
M. G. E. da Luz, C. Anteneodo, N. Crokidakis, M. Perc, \textit{Sociophysics: Social collective behavior from the physics point of view}, Chaos, Solitons $\&$ Fractals 170, 113379 (2023).








\end{thebibliography}

\end{document}